%% file: main.tex
\documentclass[conference]{IEEEtran}
\IEEEoverridecommandlockouts
\usepackage{textcomp}
\def\BibTeX{{\rm B\kern-.05em{\sc i\kern-.025em b}\kern-.08em
    T\kern-.1667em\lower.7ex\hbox{E}\kern-.125emX}}

\usepackage[T1]{fontenc}

\input{setup.tex}
\makeatletter
\def\ps@IEEEtitlepagestyle{%
  \def\@oddhead{\mycopyrightnotice}%
}
\def\mycopyrightnotice{%
  \begin{minipage}{\textwidth}
  \centering \scriptsize
  Copyright~\copyright~2022 IEEE. Personal use of this material is permitted. Permission from IEEE must be obtained for all other uses, in any current or future media, including\\reprinting/republishing this material for advertising or promotional purposes, creating new collective works, for resale or redistribution to servers or lists, or reuse of any copyrighted component of this work in other works by sending a request to pubs-permissions@ieee.org.
  \end{minipage}
}
\begin{document}
\title{Online Event Selection for Mu3e using GPUs } 
\author{\IEEEauthorblockN{Valentin Henkys}
\IEEEauthorblockA{
\textit{For the Mu3e Collaboration} \\
\textit{Institute of Computer Science} \\
\textit{Johannes Gutenberg-University} \\
Mainz, Germany \\
henkys@uni-mainz.de}
\and
\IEEEauthorblockN{Bertil Schmidt}
\IEEEauthorblockA{\textit{Institute of Computer Science} \\ 
\textit{Johannes Gutenberg-University}\\ 
Mainz, Germany \\
bertil.schmidt@uni-mainz.de}
\and
\IEEEauthorblockN{Niklaus Berger}
\IEEEauthorblockA{\textit{Institute of Nuclear Physics} \\
\textit{Cluster of Excellence PRISMA$^+$}\\
\textit{Johannes Gutenberg-University}\\
Mainz, Germany \\
niberger@uni-mainz.de
}
}

\maketitle  

\begin{abstract}

In the search for physics beyond the Standard Model the Mu3e experiment tries to observe the lepton flavor violating decay $\mu^+ \rightarrow e^+ e^- e^+$. 
By observing the decay products of $1 \cdot 10^8\mu$/s it aims to either observe the process, or set a new upper limit on its estimated branching ratio.
The high muon rates result in high data rates of $80$\,Gbps, dominated by data produced through background processes.
We present the \emph{Online Event Selection}, a three step algorithm running on the graphics processing units (GPU) of the $12$ Mu3e filter farm computers.

By using simple and fast geometric selection criteria, the algorithm first reduces the amount of possible event candidates to below $5\%$ of the initial set.
These candidates are then used to reconstruct full particle tracks, correctly reconstructing over $97\%$ of signal tracks. 
Finally a possible decay vertex is reconstructed using simple geometric considerations instead of a full reconstruction, correctly identifying over $94\%$ of signal events.

We also present a full implementation of the algorithm, fulfilling all performance requirements at the targeted muon rate and successfully reducing the data rate by a factor of $200$.
\end{abstract}

\input{chapter/introduction} %
\input{chapter/mu3e/index} %
\input{chapter/algorithm/index} %
\input{chapter/implementation/index} %

\section{Conclusion}  %

We have introduced an algorithm for the Mu3e experiment, used to filter out over $99.5\%$ of unwanted data, reducing the data rate by a factor of $>200$, doubling our target reduction rate.
It is able to correctly identify over $97\%$ of signal tracks and $>94\%$ of signal frames.
This algorithm is implemented on GPUs, achieving the stringent performance requirements of a maximum of $12$ devices on a NVIDIA Geforce GTX 1080Ti. 
Using graphics cards of newer generation, like the 2080Ti, results in spare performance. These GPUs used are already a few years old, but due to the ongoing chip shortage no newer GPUs were available to us.

Using this extra performance could allow future work to increase the efficiency even further, by possibly doing a full track reconstruction using all detector layers. 
This increase in efficiency could not only lead to a better signal track detection rate, but also improve the vertex reconstruction, due to the more precise track reconstruction.
Another point of improvement could be the vertex reconstruction, where only a rough estimate is done. 
Using the extra performance provided by newer GPUs could be invested to allow for a more precise reconstruction and a better signal detection rate.

Additionally we currently only work with fixed frames, but a particle could pass only a few layers during one frame and the rest on the next frame. 
Including these would massively increase the problem size, since multiple frames have to be processed together.

\section*{Acknowledgment}
This work was supported by the Mu3e research unit BE4876/3-1 676413 funded by DFG (German Research Foundation).

\bibliographystyle{IEEEtran}
\bibliography{library.bib}

\end{document}

%% file: setup.tex
\usepackage{graphicx}

\usepackage{hyperref}
\usepackage{color}
\usepackage{cite}

\usepackage{amsmath,amssymb}
\usepackage[capitalise]{cleveref}

\usepackage{tikz}
\usepackage{tikz-3dplot}
\usepackage{tkz-euclide}
\usepackage{pgfplots}
\usetikzlibrary{shapes.geometric,calc,angles,quotes,arrows,external,patterns,backgrounds}
\usepgfplotslibrary{groupplots,dateplot,fillbetween}
\pgfplotsset{
		compat=newest, 
}

\pgfplotstableset{col sep=comma}
\usepackage{pgfplotstable}

\newcommand*{\rzero}{25} 
\newcommand*{\rone}{31} 
\newcommand*{\rtwo}{73} 
\newcommand*{\rthree}{85} 

\newcommand*{\Phims}{\Phi_{\text{MS}}}
\newcommand*{\Thetams}{\Theta_{\text{MS}}}

\newcommand*{\cvec}[1]{\mathbf{#1}}
\usepackage{alphalph}
\makeatletter
\pgfplotsset{
    auto title/.style={
        title=(\alphalph{\pgfplots@group@current@plot})
    },
    every tick label/.append style={font=\small},
    every axis label/.append style={font=\small}
}
\makeatother

\newcommand*{\plotbars}[2]{
	\nextgroupplot[
					ybar,
					bar width=22pt,
					bar shift=0pt,
					nodes near coords,
					ylabel=#2, 
					xtick=data,
					xticklabels from table={#1}{x},
					legend style={
						},
					x=28pt,
					ymin=0,
					enlarge x limits={true, abs value=15pt},
					enlarge y limits={upper, abs value=1.5em},
					legend to name=barlegend,
					area legend,
		            xticklabel style={rotate=45, anchor=north east, inner sep=0},
				]
				\addplot+[fill opacity=1, text=black] table [x expr=\coordindex, y={mc_4seg}]{#1};
				\addplot+[fill opacity=1, text=black] table [x expr=\coordindex, y={frames}]{#1};
				\addlegendentry{True combinations};
				\addlegendentry{All combinations};
}

\usepackage{siunitx}
\usepackage{upgreek}
\usepackage{xparse}

\NewDocumentCommand{\dif}{o o}{
\IfNoValueTF{#1}
	{\mathop{}\!\mathrm{d}}
	{\frac{\mathop{}\!\mathrm{d} #1}{\mathop{}\!\mathrm{d} #2}}
}

\tikzset{
	thick line/.style={
		line width=1.3
	},
	thin line/.style={
		line width=.4
	},
}
\definecolor{electron}{rgb}{0.0, 0.32, 0.75} 
\definecolor{positron}{rgb}{1.0,0.25,0.25}
\definecolor{highlight}{rgb}{1.0,0.13,0.32}

\usepackage[linesnumbered,ruled]{algorithm2e}
\SetKwFor{ForPar}{for all}{do in parallel}{end}

\crefname{algocf}{Alg.}{Algs.}
\Crefname{algocf}{Algorithm}{Algorithms}

%% file: chapter/introduction.tex
\section{Introduction}
\label{ch:introduction}

The Standard Model of particle physics (SM) describes elementary particles and their interactions. 
One quantity originally believed to be conserved in the SM is the lepton flavor, but experiments have shown that this is not the case in the neutrino sector~\cite{Super-Kamiokande:1998kpq, SNO:2001kpb, SNO:2002tuh, nobel2015}. Flavour violating decays of charged leptons on the other
hand have never been observed and would be a clear sign of new physics beyond the Standard Model.

Modern particle physics experiments, such as \emph{Mu3e}~\cite{TDR} or ALICE~\cite{ALICE:2019ruy}, running a very high particle rates require excellent resolution and produce huge data rates. 
The upgraded ALICE experiment at LHC for example expects $3\,$TBps of raw detector data in their third run~\cite{ALICE-Gpu-usage}.
These huge amounts of data are infeasible to store and include a lot of noise and background data. 
Therefore, new systems need to be developed to process the data and identify processes of interest before storage. 
The processing needs to be done in real time and systems of \emph{Field Programmable Gate Arrays} (FPGA) and/or \emph{Graphics Processing Units} (GPU) are deployed.

For our target experiment \emph{Mu3e}, data rates of $80\,$Gbps need to be processed and filtered in real-time, possible due to a high amount of background events. Therefore a goal of reducing the data rate by a factor of at least $100$ is set.
In this work we contribute improvements on the \emph{Online Event Selection} algorithm, previously introduced by D. vom Bruch~\cite{Bruch2017a}. We reduce the complexity of the first filter step, while increasing the filter rate by another $\sim 1.5\%$.
Furthermore we implemented and tested the full algorithm using \emph{CUDA} and propose a memory layout for the incoming and outgoing data. Our implementation achieves a speedup of $2$ for our target experiment compared to the previous incomplete implementation.

\section{Related Work}
With the rise of GPGPU in recent years, GPUs have been adopted and used for many high-throughput tasks in particle physics.
One such task is the reconstruction of particle tracks, where the complexity quickly explodes due to the amount of data taken.
In~\cite{FernandezDeclara:2019ycx} Declara et al. show that GPUs are a perfect fit for tackling the problem of decoding raw detector data and reconstructing particle tracks for their high data rates of $40\,$TBits/s.

ALICE~\cite{ALICE-Gpu-usage} plans the usage of GPUs for both the offline and online phase. First GPUs are used to compress the detector data during the run. Afterwards, they propose to utilize idle times of the GPU farm, due to e.g. LHC downtimes, to perform the offline reconstruction as well.

Besides detectors designed for particle accelerators other projects, like IceCube~\cite{Hieronymus} use GPUs to work through their huge data sets. Their GPU implementation of the Pegleg algorithm, used for neutrino reconstruction, achieves an average speedup of $14$, with a maximum of over $200$.

GPUs are not only used for processing detector data, but for simulation as well. Kallenborn et al.~\cite{Kallenborn} introduced a Multi-GPU version for two common neutrino oscillation frameworks, namely \emph{Prob3++} and \emph{vSQuIDS}. The authors were able to achieve speedups of two to three orders-of-magnitude for Prob3++ and one to two orders-of-magnitude for vSQuIDS.

%% file: chapter/mu3e/index.tex
\section{The Mu3e Experiment}

The \emph{Mu3e} experiment searches for the lepton flavor violating decay $\mu^+ \rightarrow e^+ e^- e^+$, aiming to either observe it or to set a new upper limit on the branching ratio at the level of $2 \cdot 10^{-15}$ in phase I. 
This would beat the previous limit set by SINDRUM at $1 \cdot 10^{-12}$ \cite{Bellgardt:1987du} by almost three orders of magnitude. To reach this sensitivity the $\Pi E5$ beam line at the Paul-Scherrer-Institute (PSI) in Switzerland is used, delivering a muon rate of $1 \cdot 10^8\mu$/s. For phase~II the planned \emph{high intensity muon beam line} (HIMB) at PSI will be used, providing a muon rate of greater than $1 \cdot 10^9\mu$/s, increasing the possible sensitivity for Mu3e to $10^{-16}$.

\subsection{Signal and Background}
\label{sec:signal-and-background}
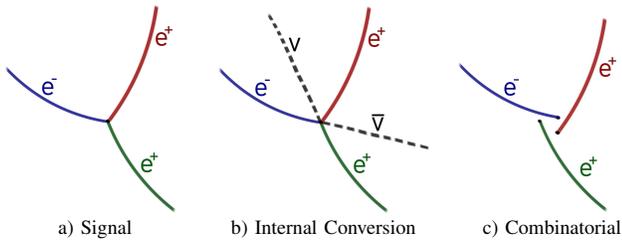
\begin{figure}[t]
	{\centering
	\resizebox{1.\columnwidth}{!}{
		\begin{tikzpicture}
			\node[inner sep=0pt] (signal) at (0,0) {\includegraphics[height=.2\textwidth]{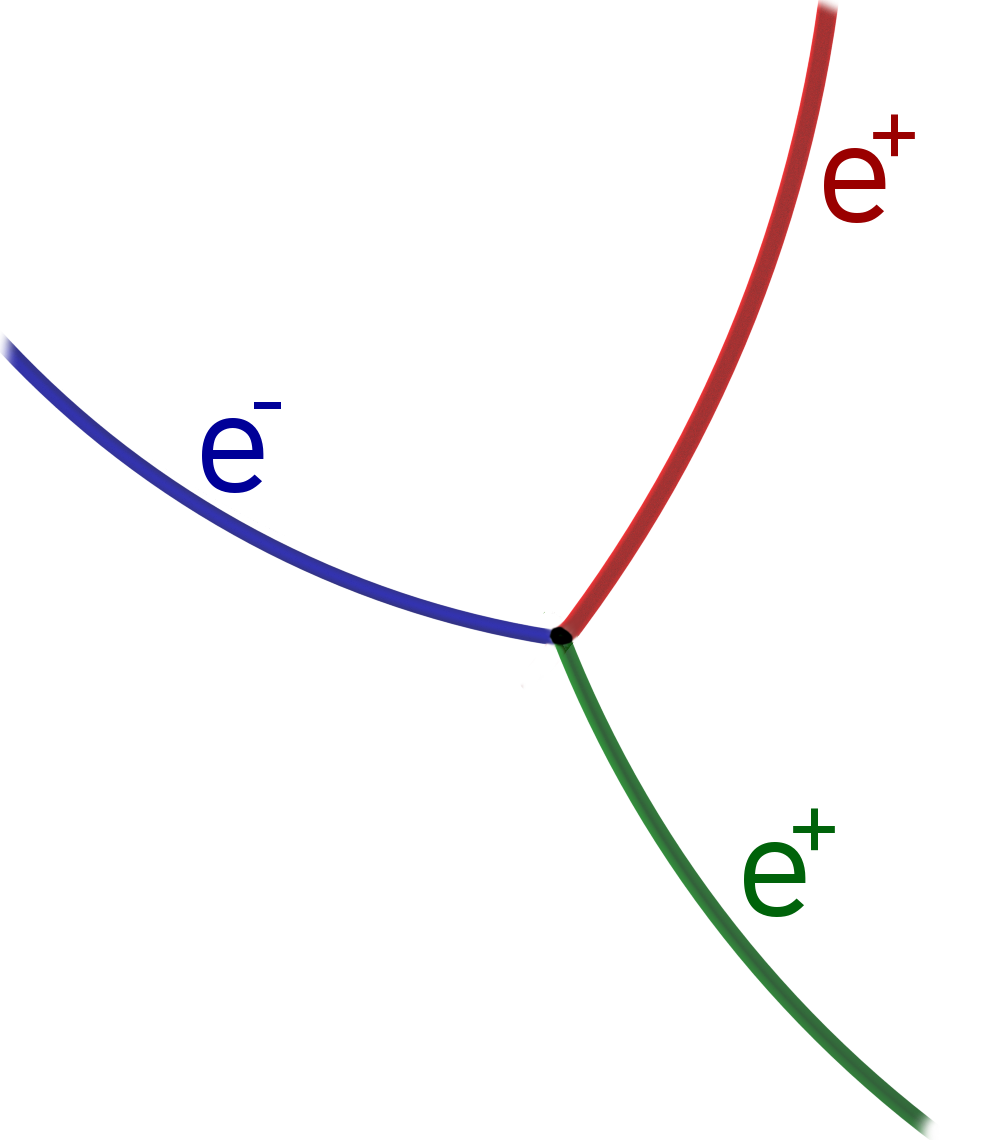}};
			\node[inner sep=0pt] (internal) at (4,0) {\includegraphics[height=.2\textwidth]{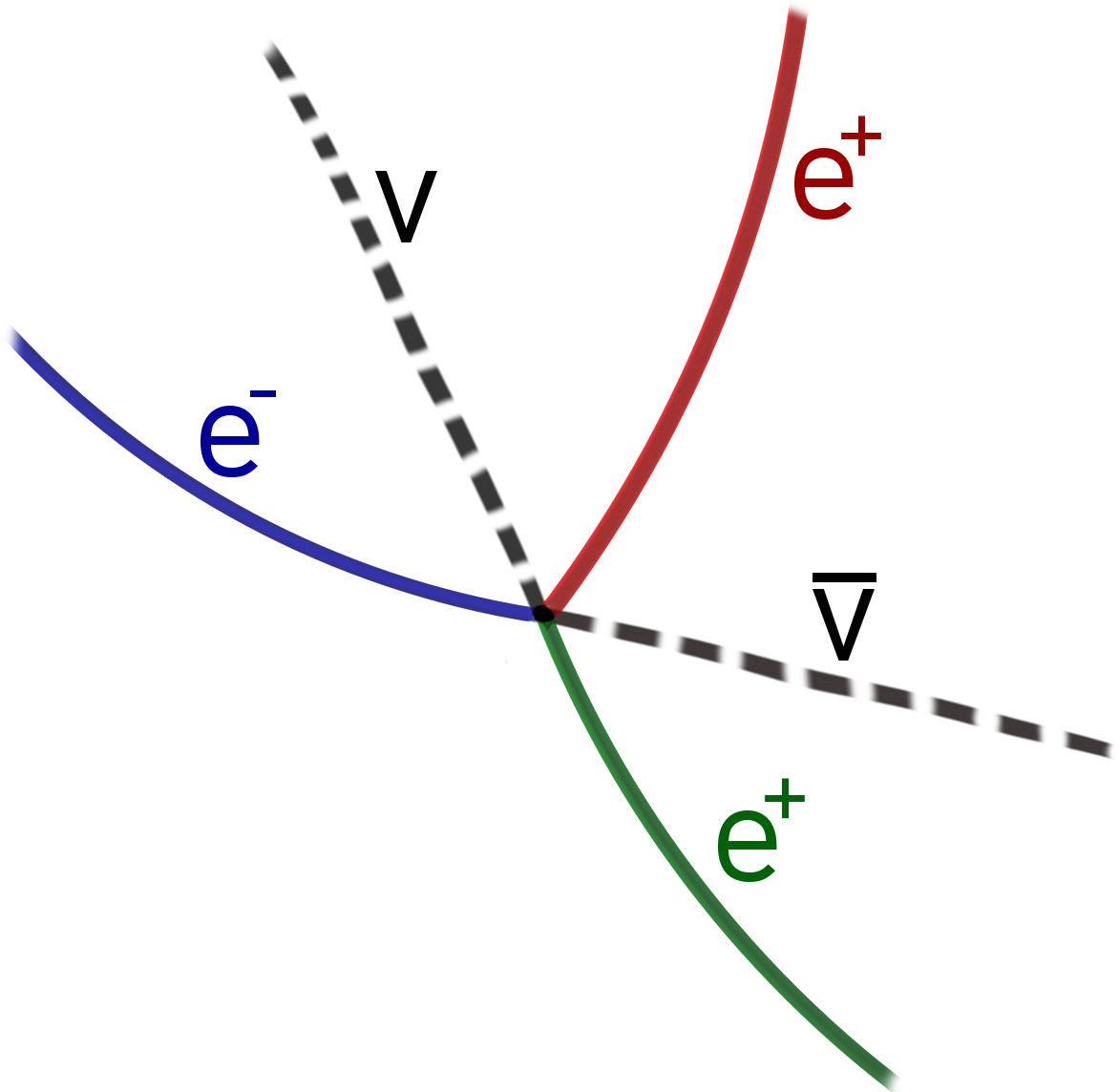}};
			\node[inner sep=0pt] (combinatorial) at (8,0) {\includegraphics[height=.2\textwidth]{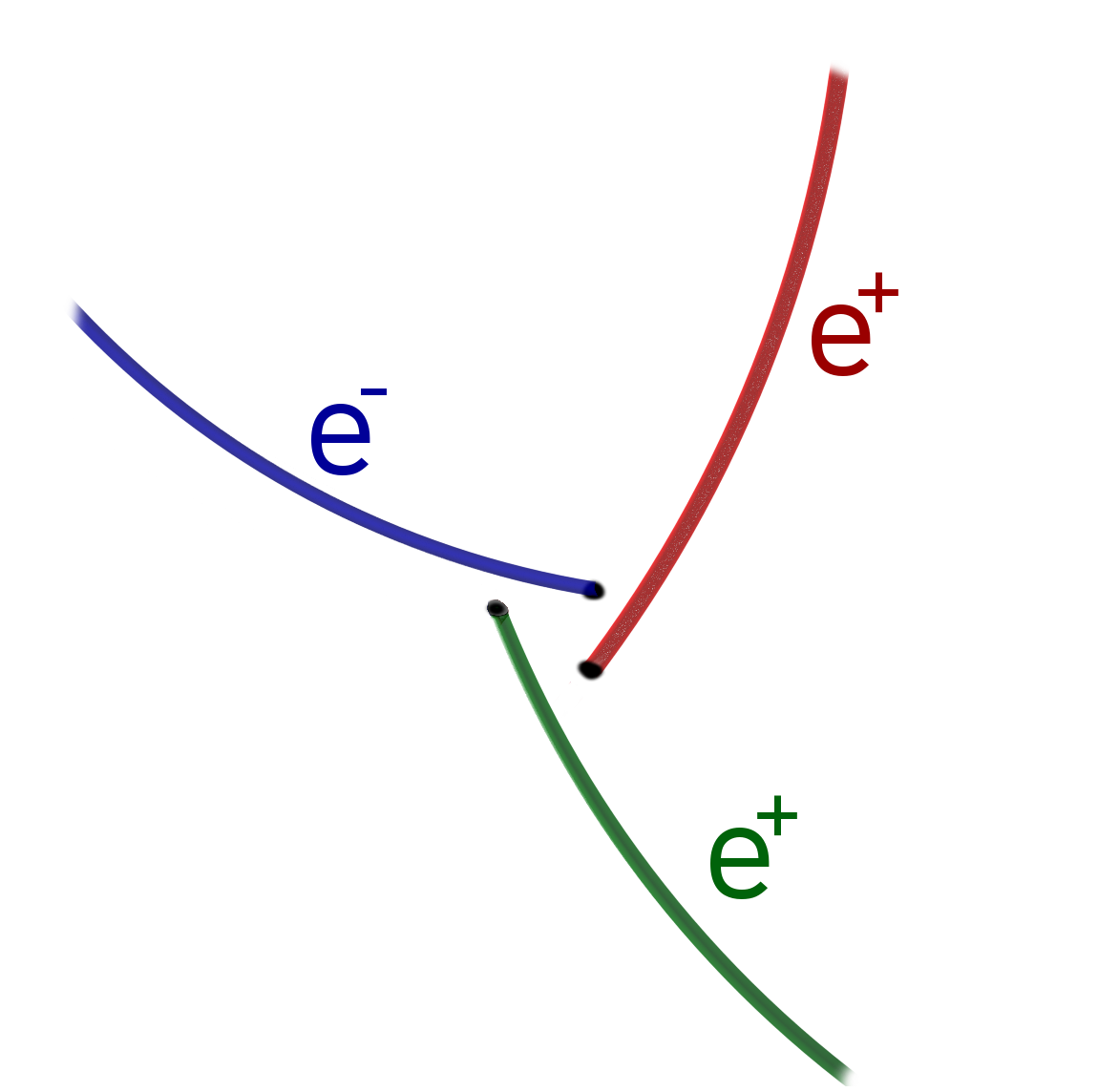}};

			\node[align = center, below=-0em of signal] {a) Signal};
			\node[align = center, below=-0em of internal] {b) Internal Conversion};
			\node[align = center, below=-0em of combinatorial] {c) Combinatorial};

		\end{tikzpicture}
	}
	}
	\caption{Schematics of the signal and background topologies. Electron tracks are in blue, positrons in red. 
		The signal a) has a clear point of origin for all tracks and no extra particles.
		In contrast, the internal conversion process b) produces two extra neutrinos, invisibly carrying away energy and momentum.
		For the combinatorial background c) no common point of origin exists for the three tracks.}
	\label{fig:signal-and-background}
\end{figure}

The signal decay $\mu^+ \rightarrow e^+ e^- e^+$ produces exactly two positrons and one electron. As the muon decays at rest, with the laws of energy and momentum conservation, the following conditions have to hold
\begin{equation}
    \sum_{i=1}^3 \cvec{p}_i = 0 \qquad \text{and} \qquad \sum_{i=1}^3 E_i = m_\mu c^2. \label{eq:conservation-laws}
\end{equation}
Furthermore, the point of origin in space and time, called event vertex, is the same for all three particles, as shown in \cref{fig:signal-and-background} a). 
These three characteristics combined can be used to discriminate a triplet of two 
positron tracks and one electron track from a signal decay from background processes.

The main background processes are radiative decays with internal conversion of the photon ($\mu^+ \rightarrow e^+ e^- e^+ \nu \bar{\nu}$, \cref{fig:signal-and-background}~b)) and combinatorial background from unrelated tracks (\cref{fig:signal-and-background}~c)).
Internal conversion results in a triplet of visible particles, created at the same vertex, but violating \cref{eq:conservation-laws}, since some momentum and energy is carried away by the neutrino and antineutrino.

The second category of background is built from superposition of different processes, creating positron and electron tracks.
The main source for positrons is the Michel decay $\mu^+ \rightarrow e^+ \nu_e \bar{\nu}_\mu$, with a branching ratio of $\sim 100\%$. 
Other sources for electron and positron tracks include, but are not limited to, photons, created by radiative decays, converting to a $e^+$/$e^-$ pair, Bhabha scattering, or wrongly reconstructed tracks, e.g.~due to detector noise. 
Triplets from these tracks either violate \cref{eq:conservation-laws} or do not originate from the same vertex.
Therefore, a good momentum and vertex resolution for the detector is needed, in order to precisely differentiate the background from the signal events.

\subsection{Mu3e Detector}
\begin{figure*}[t]
    \centering
    \includegraphics[width=.7\textwidth]{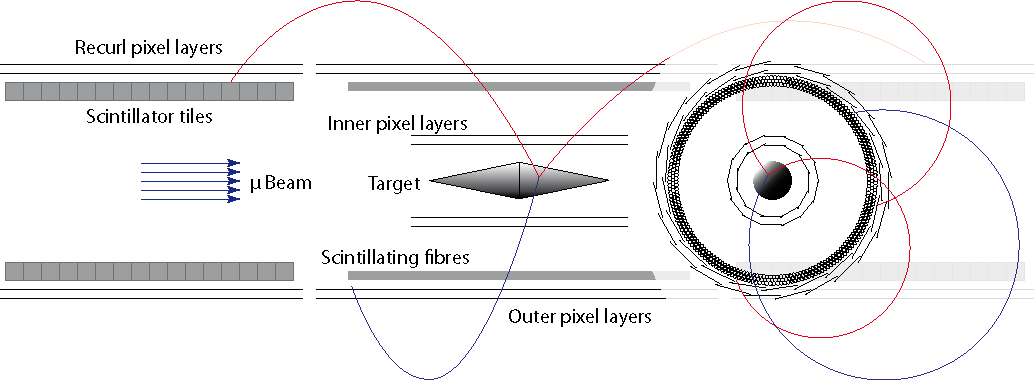}
    \caption{Schematic of the Mu3e detector, cut along the beamline, with the transverse view overlaid on the right side.~\cite{TDR}}
    \label{fig:detector}
\end{figure*}

The Mu3e detector is optimized for the detection of $\mu^+ \rightarrow e^+ e^- e^+$ and suppressing background processes. 
To achieve this goal, it is designed with a high  momentum and vertex resolution in mind.
The muon beam is stopped by a double hollow cone target, where the muons decay at rest. The whole detector sits inside a $1$T magnetic field parallel to the beam line. This results in electron and positron tracks being curved in different directions, with a curvature inversely proportional to their momentum.

Given the low momentum of the decay particles, Multiple Coulomb Scattering (MS) in the detector material dominates the momentum resolution, therefore the material budget of the detector is kept low by using ultra-thin layers of pixel detectors for tracking the particles. The pixel detectors used are \emph{high voltage monolithic active pixel sensors} (HV-MAPS) called \emph{MuPIX}~\cite{mupix}, designed with a low material budget and fast response time in mind.

Adding to the precise spatial measurements scintillating detectors are used for high precision timing measurements.
These sensors are laid out along the beam line in a cylindrical shape in three stations, with multiple layers, shown in \cref{fig:detector}.

The central station consists of four pixel and one scintillating fibre layer, used for initial track reconstruction. 
The innermost layer is wrapped closely around the target, providing a high vertex resolution~\cite{TDR}.

The outer stations, called recurl stations, are used for improving the momentum resolution by detecting particles curling back in the magnetic field using two final pixel layers and thus measuring a wider arc of the particles' helical tracks.
Being the last station used for measuring a particle, the material budget is no constraint anymore. 
Therefore thick scintillating tiles are used for time measurements, instead of thin fibres, yielding a better time resolution and dark count suppression.

%% file: chapter/algorithm/index.tex
\section{Online Event Selection}
\label{chap:algorithm}

In this section, the algorithm used for the \emph{Online Event Selection} is introduced. It is used in the Mu3e \emph{Data Acquisition System} in order to reduce the collected data rate by a factor of at least $100$ in real-time, by filtering out data from background processes. 

The concept was developed in the thesis of D.~vom Bruch~\cite{Bruch2017a} and is divided into three steps:
\begin{enumerate}
	\item \emph{Selection Cuts}: A simple geometrical filter cutting away most hit combinations before the actual track reconstruction.
	\item \emph{Track Reconstruction}: A hit triplet-based reconstruction and classification of particle tracks.
	\item \emph{Vertex Reconstruction}: A simplified reconstruction of possible event vertices. $e^+, e^+, e^-$ track combinations are examined for a possible event vertex fulfilling the signal characteristics defined in \cref{sec:signal-and-background}.
\end{enumerate}
To accommodate for the time resolution, the detected hits are bundled into timeslices, called \emph{frames}. 
Each frame is a snapshot of hits detected during a fixed timeframe. 
Due to the stringent performance requirements of real-time filtering, the problem size is reduced by only using the four central pixel layers, thus also reducing the momentum resolution.
Since the online selection process is only used to reduce the data rate as much as possible and a full offline reconstruction is done at a later stage, this reduction is acceptable and was shown to be sufficient~\cite{Bruch2017a}. 

This work presents a detailed performance study of the algorithm, resulting in improvements in background rejection as well as throughput.
Tests of the algorithm are performed and evaluated using Monte Carlo data generated with a Geant4~\cite{geant4} simulation of the full Mu3e detector. 

\input{chapter/algorithm/selection-cuts}
\input{chapter/algorithm/track-rec}
\input{chapter/algorithm/vertex-fit}
 

%% file: chapter/algorithm/selection-cuts.tex
\subsection{Selection Cuts}
\label{sec:sel-cuts}

The track reconstruction uses a triplet of hits from the first three layer as track seeds.
There are $n_0 \cdot n_1 \cdot n_2$ combinations possible, where the $n_i$ (with ${i\in\{0,1,2\}}$), are the number of detected hits in layer $i$.
Since particles may generate more than one hit and there is detector noise, only $N_\text{true} \leq \min(n_0, n_1, n_2)$ true tracks are available in the frame.
Performing a full track reconstruction on all these combinations, knowing only a fraction is real, is computationally inefficient.
As a consequence we introduce Selection Cuts, a filter consisting of four relatively simple calculations for removing over $95\%$ of all combinations, while keeping over $98\%$ of triplet combinations from true tracks.
\begin{figure}[tb]
 	\input{chapter/algorithm/figures/selection-cuts-variables}
	\label{fig:selection-cuts}
\end{figure}

The quantities used for all filter steps are shown in \cref{fig:selection-cuts}.
Since our main goal is to be fast, we simplify the detector model to a set of four concentric cylinders, such that $h_{t,k} = r_{r,i}$ being constant for all hits $k$ in layer $i$.

Using this simplification we define our filter steps based on the observation that created particles always having momenta in a narrow range.
Consequently, we can study simple geometric quantities and their behaviour on true track triplets, cutting away all possible triplet combinations falling outside of a set range.

As first filter we introduce the slope difference $\Delta z$ between the slopes of consecutive layer hits in the longitudinal plane.
The slope is defined by its angle $\lambda_{ij}$, as shown in \cref{fig:selection-cuts}
\begin{equation}
	\tan\lambda_{ij} = \frac{z_j - z_i}{r_{t,j} - r_{t,i}},\qquad i\in\{0,1\}, j=i+1.
\end{equation}
Using these the slope difference is defined as
\begin{align}
	\Delta\lambda = \tan\lambda_{12} - \tan\lambda_{01}.
	\label{eq:sel-cut-delta-lambda}
\end{align}
Moving from the longitudinal plane to the transverse plane we observe the angle $\Phi_{ij}$between hits of two consecutive layers in relation the the origin:
\begin{align}
	\cos \Phi_{ij} = \frac{\cvec{h}_{t,i} \cdot \cvec{h}_{t,j}}{r_{t,i} r_{t,j}}, \qquad i\in \{0,1\}, j = i+1.
	\label{eq:sel-cut-phi}
\end{align}
As final filter we use the transverse radius of the circle going through all three hits
\begin{equation}
	r_{t,c} = \frac{d_{01} d_{12} d_{20}}{2 [(\cvec{h}_0 - \cvec{h}_1) \times (\cvec{h}_2 - \cvec{h}_1)]_z},
	\label{eq:rt-cut}
\end{equation}
with $d_{ij} = \left| \cvec{h}_i - \cvec{h}_j \right|$~\cite{Berger2016a}. 
This last step is computationally complex, especially compared to the first three.
But since we use it as last step in the algorithm it is only used on a fraction of the initial combinations, helping to cut off about $1$ last percent. 
For all kept triplets the result can be stored and reused for the track reconstruction, described in the next section.

Assembling all filter steps and testing them on simulated data, as shown in \cref{fig:selection-cuts-versions}, results in over $95\%$ of possible combinations being cut away, while retaining over $98.5\%$ of true triplet combinations. 
We sorted the filters by impact, where the first step removes already over $80\%$ of possible combinations.
\begin{figure}[tb]
 	\input{chapter/algorithm/figures/selection-cuts-full}
	\label{fig:selection-cuts-versions}
\end{figure}
The previous version of the Selection Cuts discussed in \cite{Bruch2017a} cuts away only $93.5\%$ of combinations (whilst keeping $1\%$ additional true combinations), while using one more filter step and with all filter steps being computationally more complex.
While keeping more true combinations is favorable, this difference does not make a difference in accuracy for the full algorithm.
In contrast cutting away one more percent from all combinations does result in a measurable performance boost, due to the low computational complexity of this step compared to the track reconstruction.

%% file: chapter/algorithm/figures/selection-cuts-variables.tex
\centering
\begin{tikzpicture}[inner frame sep=0, outer frame sep=0]
\begin{groupplot}[group style={	group size=2 by 1,
										vertical sep=3em,
										horizontal sep=0.5em,
										every plot/.style={
												ymin=-100,
												ymax=100,
												axis equal,
											},
										},
								]
	\nextgroupplot[xmin=-100,
									xmax=100, 
									width=.5\columnwidth, 
									height=.5\columnwidth,
									axis x line=bottom,
									axis y line=left,
									xlabel={x [mm]},
									ylabel={y [mm]},
							    y label style={at={(axis description cs:-0.07,.5)},anchor=south},
									ticklabel style={font=\tiny},
						]
		\coordinate (C) at (axis cs: 0, 0);
		\coordinate (R0) at (axis cs: 0, \rzero);
		\coordinate (R1) at (axis cs: 0, \rone);
		\coordinate (R2) at (axis cs: 0, \rtwo);
		\coordinate (R3) at (axis cs: 0, \rthree);

		\coordinate (p0t) at (axis cs: -10.260084,20.74582);
		\coordinate (p1t) at (axis cs: 16.795656,24.376202);
		\coordinate (p2t) at (axis cs: 66.512955,-27.633915);
		\addplot[highlight, only marks, mark=x, mark size=4pt, ultra thick] coordinates {
				(0-10.260084, 20.74582)
				( 16.795656, 24.376202)
				(66.512955, -27.633915)
			};
	\nextgroupplot[xmin=-200,
					xmax=200, 
					width=.7\columnwidth, 
					height=.5\columnwidth, 
					axis x line=bottom,
					axis y line=left,
					xlabel={z [mm]},
					ylabel={},
					yticklabels={,,,,},
					enlarge y limits={false},
					enlarge x limits={false},
					 y label style={at={(axis description cs:-0.05,.5)},anchor=south},
					ticklabel style={font=\tiny},
									] 
		\coordinate (z00) at (axis cs: -60, \rzero);
		\coordinate (z01) at (axis cs: 60, \rzero);
		\coordinate (z02) at (axis cs: -60, -\rzero);
		\coordinate (z03) at (axis cs: 60, -\rzero);

		\coordinate (z10) at (axis cs: -60, \rone);
		\coordinate (z11) at (axis cs: 60, \rone);
		\coordinate (z12) at (axis cs: -60, -\rone);
		\coordinate (z13) at (axis cs: 60, -\rone);

		\coordinate (z20) at (axis cs: -170, \rtwo);
		\coordinate (z21) at (axis cs: 170, \rtwo);
		\coordinate (z22) at (axis cs: -170, -\rtwo);
		\coordinate (z23) at (axis cs: 170, -\rtwo);

		\coordinate (z30) at (axis cs: -180, \rthree);
		\coordinate (z31) at (axis cs: 180, \rthree);
		\coordinate (z32) at (axis cs: -180, -\rthree);
		\coordinate (z33) at (axis cs: 180, -\rthree);

		\coordinate (targetTop) at (axis cs: 0,19);
		\coordinate (targetBottom) at (axis cs: 0,-19);
		\coordinate (targetLeft) at (axis cs: -50,0);
		\coordinate (targetRight) at (axis cs: 50,0);

		\coordinate (p0z) at (axis cs: 12, \rzero);
		\coordinate (p1z) at (axis cs: -15, \rone);
		\coordinate (p2z) at (axis cs: -100, \rtwo);

		\coordinate (p1z0) at (axis cs: -30, \rone);

		\addplot[highlight, only marks, mark=x, mark size=4pt, ultra thick] coordinates {
				(10, \rzero)
				(-10, \rone)
				(-100, \rtwo)
			};
\end{groupplot}
\begin{scope}[thick, on background layer]
    \foreach \n in {8, 10, 24, 28} {
        \pgfmathsetmacro\angle{360 / \n * -1};
        \ifthenelse{\equal{\n}{8}}
        	{
		        \pgfmathsetmacro\startAngle{\angle};
		        \coordinate (A) at (R0);
	        }
        	{
        		\ifthenelse{\equal{\n}{10}}
	        		{
				        \pgfmathsetmacro\startAngle{\angle * -.5};
	        			\coordinate(A) at (R1);
	        		}
	        		{
	        			\ifthenelse{\equal{\n}{24}}
	        				{
						        \pgfmathsetmacro\startAngle{\angle};
	        					\coordinate (A) at (R2);
	        				}
	        				{
						        \pgfmathsetmacro\startAngle{\angle * -.5};
	        					\coordinate (A) at (R3);
	        				};
	        		};
	        };
        \coordinate (A) at ([rotate around={\startAngle:(C)}]A);
        \foreach \i [evaluate=\i] in {0,1,...,\n} {
            \wlog{tmp: \i};
            \coordinate (B) at ([rotate around={\angle:(C)}]A);
            \draw[gray] (A) -- (B);
            \coordinate (A) at (B);
        }
    }

		\draw[gray] (z00) -- (z01);
		\draw[gray] (z02) -- (z03);
		\draw[gray] (z10) -- (z11);
		\draw[gray] (z12) -- (z13);
		\draw[gray] (z20) -- (z21);
		\draw[gray] (z22) -- (z23);
		\draw[gray] (z30) -- (z31);
		\draw[gray] (z32) -- (z33);

		\tkzDrawPolygon[fill=black!10,thin](targetLeft,targetTop,targetRight,targetBottom)

	\end{scope}

		\begin{scope}
			\tkzDrawLines[add=0 and 0](p0t,C C,p1t);
			\tkzDrawLines[add=0 and 0](p2t,C);

			\tkzMarkAngle[dashed, size=.35, mark=none](p1t,C,p0t);

			\tkzDefPointOnLine[pos=.5](p0t,p1t) \tkzGetPoint{tmp1}
			\tkzDefPointOnLine[pos=.5](tmp1,C) \tkzGetPoint{tmp2};
			\tkzDefPointOnCircle[angle=110, center=C, radius=.5] \tkzGetPoint{tmp3};

			\draw[stealth-] (tmp2) to[bend right] (tmp3) node[above left, outer sep=0, inner sep=0] {$\Phi_{01}$};

			\tkzDrawLines[dashed, add=1 and 0](p1t,C);
			\tkzMarkAngle[dashed, size=.95, mark=none](p2t,C,p1t);
			\tkzLabelAngle[pos=.7](p2t,C,p1t){$\Phi_{12}$};


			\draw[decorate,decoration={brace,amplitude=5pt,raise=0.5pt},yshift=0pt] (p2t) -- (C) node [midway,yshift=-9pt]{$h_{t,2}$};

			\tkzMarkAngle[dashed, arc=lll, size=1.1, mark=none](p2z,p1z,p1z0);
			\tkzLabelAngle[pos=.8](p2z,p1z,p1z0){$\lambda_{12}$};
			\tkzDrawLine[dashed, add=0 and 0](p1z,p2z);

		\end{scope}
y\end{tikzpicture}
    \vspace*{-.7cm}
	\caption{Sketch of Selection Cuts and the geometric quantities used. The red crosses are hits from a triplet combination of the first three detector layers.~Based on \cite{Bruch2017a}}

%% file: chapter/algorithm/figures/selection-cuts-full.tex
\begin{center}
\begin{tikzpicture}
	\begin{groupplot}[
		group style={group size=1 by 1, horizontal sep=2em},
		width=\columnwidth-3em,
		height=4cm, 
		]
		\plotbars{plots/sel_cuts/bars.csv}{$\text{combinations kept [}\%\text{]}$}; 
		\coordinate (c1) at (rel axis cs:1,0);
		\coordinate (c2) at (rel axis cs:1,1);

	\end{groupplot}
		\coordinate (c3) at ($(c1)!.5!(c2)$);
    \node[right, scale=.75, transform shape] at (c3 |- current bounding box.east) {\pgfplotslegendfromname{barlegend}};
\end{tikzpicture}
\end{center}

	\vspace*{-.3cm}
\caption{The amount of combinations kept after applying each filter step of the Selection Cuts one after another. The first filter already removes over $80\%$ from all triplet combinations, keeping almost all true triplet combinations. Each consecutive step almost halves the previous amount of triplet combinations, with the last step only removing one final percent.}

%% file: chapter/algorithm/track-rec.tex
\subsection{Track Reconstruction}
\label{sec:track-rec}

Starting from the previously selected triplets, full particle tracks need to be reconstructed. 
For the reconstruction, the \emph{Triplet Fit}~\cite{Kozlinskiy2017,Berger2016a} is used. 
This algorithm provides an analytical solution, without the need for an iterative approach, like in the case of the Kalman-Filter~\cite{Ai2021}, or expensive matrix inversions as in the Broken Lines Fit~\cite{Kiehn,Blobel2006}. 
The Triplet Fit is designed for Multiple Scattering (MS) dominated environments, by fitting a helix to hit triplets, where MS is occurring in the central layer.

\subsubsection{Single Triplet Fit}
Hit triplets used for the Triplet Fit represent hits in consecutive layers, assumed to have negligible spatial uncertainties.
MS in the central layer results in a kink in the trajectory.
Assuming no momentum loss and thus a constant curvature $\kappa$, this trajectory change is described by a sudden change in the track angles $\Phi$ and $\lambda$, called $\Phi_{MS}$ and $\Theta_{MS}$. An example trajectory including the variables used in this section is shown in \cref{fig:triplet-fit}.
The distribution of the scattering angles $\Phims$ and $\Thetams$ is assumed to be Gaussian with a mean of zero and variances given by MS theory $\sigma^2_\Theta = \sigma^2_{MS}$ and $\sigma^2_\Phi = \tfrac{\sigma^2_{MS}}{\sin^{2} \lambda}$~\cite{Highland1975,Lynch1991}.
Furthermore, assuming no momentum loss and the correlation $\kappa \propto \tfrac{1}{p}$ we search for the track minimizing the objective function~\cite{Berger2016a}
\begin{equation}
	\chi^2 (\kappa) = \frac{\Phims (\kappa)^2}{\sigma_\Phi^2} + \frac{\Thetams (\kappa)^2}{\sigma^2_\Theta}.
	\label{eq:objective-function}
\end{equation}

The minimization problem is non-linear, but we can linearize it around some known solution. 
The scattering angles are expected to be small, therefore we can use the solution where $\Phims = 0$.
This solution forms a circle in the transverse plane, and all parameters can be rewritten with a dependence on the circle radius $r_{t,c}$. 
Now we can solve the problem using a first order Taylor expansion around the circle solution~\cite{Berger2016a}.

\begin{figure}[t!]
    \input{chapter/algorithm/figures/triplet-fit}
    \label{fig:triplet-fit}
\end{figure}

\subsubsection{Triplets Fit}
\label{sec:triplet-fit}

Incorporating more than three hits in the algorithm for a full track fit requires to accommodate for multiple triplets in the objective function. 
Each consecutive triplet uses the last two hits of the previous triplet as first two and adds a hit from the next layer as third hit.
The goal is to find one global curvature for all triplet combinations minimising the MS angles for each triplet.
This results in the new, global objective function~\cite{Bruch2017a}
\begin{equation}
	\chi^2_\text{global} (\kappa) = \sum_t^{n_\text{triplets}} \chi^2_t(\kappa) .
	\label{eq:objective-function-global}
\end{equation}

The scattering angles $\Phi_{MS,t}$ and $\Theta_{MS,t}$ for each triplet are independent of the other triplets. Consequently, each $\chi^2_t$ is minimised individually, and the global curvature is defined as the weighted average~\cite{Bruch2017a}
\begin{equation}
	\bar{\kappa} = \frac{ \sum_t^{n_\text{triplets}} \frac{\kappa_t}{\sigma^2_{\kappa,t}} }
	{\sum_t^{n_\text{triplets}} \frac{1}{\sigma^2_{\kappa,t}}}.
\end{equation}
Adding more hits, and thus triplets increases the momentum resolution~\cite{Berger2016a}. 
For the Online Event Selection, only the first four layers are used, thus two triplets need to be fit. 
First the helix for the combinations chosen by the Selection Cuts is fit. 
Using this preliminary helix, the hit position in the fourth layer is estimated. 
The estimated point is used for finding the closest fourth layer hit, which is then used to build the second triplet and perform a fit.
Finally, only tracks with a $\chi^2$ error of smaller than $32$ are kept, resulting in $94\%$ of true tracks being kept. 

For the tracks kept, the track parameters are calculated, and the tracks are classified as electrons and positrons, depending on the sign of their global curvature $\kappa$.

%% file: chapter/algorithm/figures/triplet-fit.tex
\begin{tikzpicture}

	\def\dotSize{2pt}
	\coordinate[label=left:$h_0$] (h0) at (0,0);
	\coordinate[label=above:$h_1$] (h1) at (3,1.5);
	\coordinate[label=above:$h_2$] (h2) at (6,0.8);

	\coordinate[] (h2_ghost) at (4.5,0.6);
	\coordinate[] (h0_ghost) at (2.8,1.4);

	\fill (h0) circle (\dotSize);
	\fill (h1) circle (\dotSize);
	\fill (h2) circle (\dotSize);

	\draw[dashed] (h0) -- node[below,midway] (d01) {} (h1);
	\draw[dashed] (h1) -- node[below,midway] (d12) {} (h2);

	\tkzCircumCenter(h0,h1,h2_ghost)\tkzGetPoint{c01};
	\node[label=left:$c_{01}$] at (c01) {};
	\fill (c01)  circle (\dotSize);
	\tkzDrawArc[color=electron](c01,h1)(h0);

	\tkzCircumCenter(h0_ghost,h1,h2)\tkzGetPoint{c12}
	\fill (c12)  circle (\dotSize);
	\node[label=right:$c_{12}$] at (c12) {};
	\tkzDrawArc[color=electron](c12,h2)(h1);
	\tkzLabelCircle[above=0](c12,h2)(15){$s_{12}$};
	\tkzLabelCircle[above=0](c01,h1)(35){$s_{01}$};

	\draw (h0) -- (c01) -- (h1);
	\tkzLabelLine[pos=.5,below](h0,c01){$r_{t,01}$};
	\draw (h1) -- (c12) -- (h2);
	\tkzLabelLine[pos=.5,right](h2,c12){$r_{t,12}$};

	\tkzMarkAngle[arc=lll, size=1, mark=none](h1,c01,h0)
	\tkzLabelAngle[pos=0.6](h1,c01,h0){$\Phi_{01}$}

	\tkzMarkAngle[arc=lll, size=1, mark=none](h2,c12,h1)
	\tkzLabelAngle[pos=0.6](h2,c12,h1){$\Phi_{12}$}



	\tkzMarkAngle[arc=lll, size=1.5, mark=none](c01,h1,c12)
	\tkzLabelAngle[pos=1.1](c01,h1,c12){$\Phims$}

	\coordinate[] (axis_0) at (current bounding box.south west);
	\coordinate[] (axis_y_max) at (current bounding box.north west);
	\draw[draw, -stealth, ultra thick] (axis_0) -- ++(.5,0) node[below] {$x$};
	\draw[draw, -stealth, ultra thick] (axis_0) -- ++(0,.5) node[left] {$y$};
\end{tikzpicture}
\centering
\begin{tikzpicture} 
	\def\dotSize{2pt}

	\coordinate[label=above:$h_0$] (h0) at (0,0);
	\coordinate[label=above:$h_1$] (h1) at (1,.75);
	\coordinate[label=above:$h_2$] (h2) at (3.5,1.25);

	\tkzDrawLine[dashed, add=0 and 0.6](h0,h1)
	\tkzDrawLine[color=electron, add=0 and 0](h1,h2)
	\tkzDrawLine[color=electron, add=0 and 0](h0,h1)
	
	\tkzDefPointOnLine[pos=1.2](h0,h1)
	\tkzGetPoint{tmp}
	\tkzLabelAngle[pos=1.2](h2,h1,tmp){$\Thetams$}
	\tkzMarkAngle[size=0.8, mark=none](h2,h1,tmp)

	\coordinate (x_axis) at (1,0);
	\tkzDrawLine[add=0 and 0](h0,x_axis)
	\tkzLabelLine[pos=.5,below](h0,x_axis){$z_{01}$}
	\tkzDrawLine[dashed, add=0 and 0](h1,x_axis)

	\coordinate (x_axis) at (3.5,.75);
	\tkzDrawLine[add=0 and 0](h1,x_axis)
	\tkzLabelLine[pos=.5,below](h1,x_axis){$z_{12}$}
	\tkzDrawLine[dashed, add=0 and 0](h2,x_axis)

	\fill (h0) circle (\dotSize);
	\fill (h1) circle (\dotSize);
	\fill (h2) circle (\dotSize);

	\coordinate[] (axis_0) at (current bounding box.south west);
	\coordinate[] (axis_y_max) at (current bounding box.north west);
	\draw[draw, -stealth, ultra thick] (axis_0) -- ++(.5,0) node[below] {$z$};
	\draw[draw, -stealth, ultra thick] (axis_0) -- ++(0,.5) node[left] {$s$};
\end{tikzpicture}

	\vspace*{-.3cm}
\caption{Sketch of a reconstructed track going through three hits $h_0, h_1, h_2$ with a kink from MS in the central hit. ~Based on \cite{Berger2016a}.}

%% file: chapter/algorithm/vertex-fit.tex
\subsection{Vertex Fit}
\label{sec:vertex-fit}

Having possible tracks for a frame reconstructed, we now want to check if the frame looks as if a signal event occurred.
Keeping the computational costs low, we define our goal to not have an exact reconstruction, but rather having a rough estimate of a possible spatial event vertex.
Each track triplet consisting of two positron tracks and one electron track is analyzed and checked for a possible event signature, which is defined in \cref{sec:signal-and-background}.
The concept behind the algorithm is to first reduce the problem complexity by finding a possible event vertex in the transverse plane and only if such a vertex was found, to project it back into the third dimension.
Back in three dimensions the possible vertex is checked for signal compatibility. 
If a compatible vertex is found, the frame will be kept for storage and offline analysis.

First a track triplet's total energy is calculated. If it differs too much from the muon's rest energy, it violates energy conservation and is discarded.
Otherwise, the tracks are processed and checked for a possible vertex.

\subsubsection{Finding Possible Event Vertices}
Next the tracks are projected onto circles in the transverse plane, which are defined by their center $\cvec{c}_i$ and radii $r_{t,i} = \tfrac{1}{\kappa_{t,i}}$.

Using the circle representation, all circle-circle intersections between all tracks are determined. 
If two circles do not intersect, the track triplet is skipped.
Otherwise, the points of intersection $\cvec{p}_{0/1}$ have to be calculated.

All intersections too far away from the target, represented by a disk with a radius of $19$mm, are dismissed.
Using all combinations of intersection triplets, consisting of one intersection per track pair, a possible $2$D event vertex $\mu_t$ is estimated.
It is defined as weighted mean $\mu_t$ of the intersection points $\cvec{p}_i, i\in\{0,1,2\}$ weighted by their uncertainties~$\sigma_i^2$~\cite{Bruch2017a}
\begin{equation}
	\cvec{\mu}_t = \frac{\sum_{i=0}^3 \frac{\cvec{p}_i}{\sigma_i^2}}{\sum_{i=0}^3 \frac{1}{\sigma^2_i}}.
\end{equation}
Since the first layer is wrapped closely around the target, the intersection points may be relatively close to the detector layer, where the pixel error dominates over the MS induced error.
Therefore, the uncertainties for each point are defined by both, the pixels spatial resolution $\sigma_\text{Pixel}^2$ and the tracks $\sigma_\text{MS}^2$ error~\cite{Bruch2017a}
\begin{equation}
	\sigma^2_i = \sigma_{MS}^2 \cdot s_i^2 + \sigma_\text{Pixel},
	\label{eq:track-error}
\end{equation}
with $s_i$ as the path length along the circle from the measured detector hit in layer~$0$.

\begin{figure}[t]
	\input{chapter/algorithm/figures/pca}
	\label{fig:pca}
\end{figure}

Next we want to find the point corresponding to the event vertex. 
This point is defined as the point of closest approach $\cvec{p}_{ca,i}$ to the mean position on each track, calculated as shown in \cref{fig:pca}. 

This circle point is now projected back onto the $3$D helix track. 
To achieve this, the angle $\Delta\Phi$ traveled from the initial hit $\cvec{h}_{i,t}$ to $\cvec{p}_{t,ca,i}$ is calculated and used to find the corresponding $z$-position on the helix
\begin{equation}
	z_i = \cvec{h}_{i,z} - \frac{\Delta\Phi \sin \lambda_{01}}{\kappa}.
\end{equation}

Again the error is calculated, following \cref{eq:track-error}. 
Using all three $z$-positions the mean $\mu_{z}$ is calculated and used as $z$-coordinate for our estimated event vertex.
The possible signal vertex position $\mu$ for this triplet combination is defined using $\mu_{z}$ and $\mu_t$.

\subsubsection{Signal Estimation}

Using all points from the previous section, we now calculate the distance from each closest helix point to the vertex position $\mu$ and its error~\cite{Bruch2017a}
\begin{equation}
	\chi^2 = \sum^3_{i=0} \frac{\mu_i}{\sigma^2_i}.
\end{equation}

Among all vertices found from all track triplets we are using only the one with the smallest $\chi^2$ error. 
If the smallest error is larger than a threshold, it is discarded as well.
Otherwise, we test the proximity to the target surface for the chosen vertex. Events to far away are discarded.
Otherwise, the total momentum of all tracks at the points of closest approach is estimated.
If the vectorial momentum is too high, it violates momentum conservation as defined in \cref{sec:signal-and-background} and the frame is discarded.
In all other cases the frame is kept.

%% file: chapter/algorithm/figures/pca.tex
\centering
	\begin{tikzpicture}[scale=3,thick]
		\def\rt{1}
		\tkzDefPoints{0/0/c_0,.5/1/c_1,1.5/.35/c_2}
		\tkzDrawPoints[shape=cross out](c_0,c_1,c_2)
		\tkzLabelPoints[below left](c_0);
		\tkzLabelPoints[above](c_1);
		\tkzLabelPoints[right](c_2);
		\tkzDefPointOnCircle[angle=30,center=c_0,radius=\rt]
		\tkzGetPoint{r_0}
		\tkzDefPointOnCircle[angle=30,center=c_0,radius=\rt]
		\tkzGetPoint{r_1}
		\tkzDefPointOnCircle[angle=50,center=c_0,radius=\rt]
		\tkzGetPoint{r_2}

		\begin{scope}
			\tkzDefPointOnCircle[angle=-10,center=c_0,radius=\rt]
				\tkzGetPoint{A};
			\tkzDefPointOnCircle[angle=110,center=c_0,radius=\rt]
				\tkzGetPoint{B};
			\tkzDrawArc[color=electron,thick](c_0,A)(B);
		\end{scope}
		\begin{scope}
			\tkzCalcLength[cm](c_1,r_1) \tkzGetLength{C};
			\tkzDrawArc[R,color=positron,thick](c_1,\C)(170,370);

			\tkzDefPointOnCircle[angle=230,radius=\C,center=c_1] \tkzGetPoint{P};
			\tkzDrawLine[add=0 and 0](c_1,P);
			\tkzLabelLine[pos=.5,left](c_1,P){$r_{t,i}$};
		\end{scope}
		\begin{scope}
			\tkzCalcLength[cm](c_2,r_2) \tkzGetLength{C};
			\tkzDrawArc[R,color=positron,thick](c_2,\C)(130,210);
		\end{scope}
		\tkzInterCC(c_1,r_1)(c_2,r_2)\tkzGetPoints{r_12}{r_21}
		\tkzDrawPoints[purple!70](r_0,r_2,r_21)

		\tkzDefTriangleCenter[mittenpunkt](r_0,r_2,r_21)
			\tkzGetPoint{m};
		\tkzDrawPoint[shape=cross out](m);
		\tkzLabelPoint[right](m){$\mu_t$};

		\foreach \pt in {0,1,2} {
			\tkzDefPointOnLine[pos=2](c_\pt,m) \tkzGetPoint{m_\pt}
			\tkzDefPointOnLine[pos=-22](c_\pt,m) \tkzGetPoint{tmp_\pt}
			\tkzInterLC(tmp_\pt,m_\pt)(c_\pt,r_\pt)		\tkzGetPoints{tmp}{ca_\pt}
			\tkzDrawPoint(ca_\pt)
		}
		\tkzDrawLine[add=0 and 0, -stealth](c_1,m);
		\tkzLabelLine[pos=.5,left](c_1,m){$\cvec{d}$};

		\tkzLabelPoint[left=0.5cm of ca_0](ca_0){$\cvec{p}_{t,ca,2}$}
		\tkzLabelPoint[below right=0cm and -.1cm of ca_1](ca_1){$\cvec{p}_{t,ca,1}$}
		\tkzLabelPoint[above right=.1cm and .5cm of ca_2](ca_2){$\cvec{p}_{t,ca,0}$}

	\coordinate[] (axis_0) at (current bounding box.south west);
	\coordinate[] (axis_y_max) at (current bounding box.north west);
	\draw[draw, -stealth, very thick] (axis_0) -- ++(.15,0) node[below] {$x$};
	\draw[draw, -stealth, very thick] (axis_0) -- ++(0,.15) node[left] {$y$};
	\end{tikzpicture}

	\vspace*{-.3cm}
	\caption{Sketch showing how to estimate the transverse point for a event vertex and the points of closest approach for a track triplet.~Based on \cite{Bruch2017a}.}

%% file: chapter/implementation/index.tex
\section{Implementation}
\label{ch:implementation}

We implemented the Online Event Selection using \emph{CUDA}~\cite{cuda} for one GPU, where the data is supplied by the computers FPGA.
The computers in the Mu3e filter farm are daisy chained together, with each computer working independently on their individual batch of consecutive frames. 
Therefore, the implementation does not need to accommodate for Multi-GPU setups and will scale linearly with the amount of computers used.

\subsection{Memory Layout}
As the detector detects hits the data is streamed to the filter farm, where the data is prepared for the GPUs. 
We propose a memory layout optimised for GPU access patterns, while allowing the FPGA to efficiently fill it with incoming data. 
We start by collecting multiple frames in chunks, allowing for one chunk to be transferred to and processed by the GPU, while the next one is filled with incoming data.

Due to the highly varying size of one frame, the chunks are fixed in size with varying amount of frames. 
They consist of two arrays, and one integer.
The first array starts at the beginning of the chunk, collecting the frame's hit data, sorted by frame and layer. 
Each hit is stored as a $(x,y,z)$ vector. 
As last element of the chunk an integer is placed, counting the number of frames stored in this chunk. Starting from the second to last position the second array collects the pointer to each frame and frames layer start position. 

A chunk's creation is stopped and a new one is created, when inserting the new frame would lead to an overflow. 
This layout allows for simple block wise transfer to the GPU, while keeping memory overhead low and the amount of calculations needed for frame metadata (like hits per layer) to a minimum.

\subsection{Parallelization}

\begin{algorithm}[t!]
\DontPrintSemicolon
    \tcp{Distribute frames over CUDA blocks}
    \ForPar{frame $\in$ frames}{
        load\_into\_shared\_memory(frame)\;
        \_\_syncthreads()\;
        selection\_cuts(frame) \;
        \_\_syncthreads()\;
        track\_reconstruction(frame) \;
        \_\_syncthreads()\;
        vertex\_fit(frame) \;
        \_\_syncthreads()\;
        store\_frame(frame)
    }
    \caption{Main kernel}
    \label{alg:main}
\end{algorithm}

In this subsection we will describe the schema used to parallelize the algorithm. 
Each frame's selection process is independent from one another, therefore we parallelize over the frames, as shown in \cref{alg:main}.
One CUDA block processes a chunk of frames, frame by frame. 
Even though number of hits per frame vary highly and therefore each frame has a different computational cost, different distribution schemes did not yield any noticeable performance difference. Thus, we chose a cyclic distribution.

\begin{algorithm}
\DontPrintSemicolon
    \tcp{Distribute across threads}
    \ForPar{hit\_combinations(frame)}{
        \If{$test\_\Delta\lambda (hit_0, hit_1, hit_2)$}{
            \If{$test\_\Phi_{01}(hit_0, hit_1)$} {
                \If{$test\_\Phi_{02}(hit_1, hit_2)$} {
                    \If{$test\_r_{t,c}(hit_0, hit_1, hit_2)$} {
                        \If{num\_cuts $<$ cuts\_max} {
                            save\_cut($hit_0, hit_1, hit2$)\;
                            $num\_cuts++$
                        }
                    }
                }
            }
        }
    }
    \caption{Selection Cuts}
    \label{alg:sel-cuts}
\end{algorithm}
When processing a frame, first the hit data is loaded into shared memory, for faster access. 
Since each step of the algorithm is dependent on the previous one, we first start by performing the Selection Cuts, shown in \cref{alg:sel-cuts}.
Each triplet calculation is independent from the others, so each thread performs the cuts on a chunk of triplet combinations.
Since each filter step reduces the amount of triplets by a big chunk and is computationally simple, thread divergence is kept low and short. 
Selected triplets are then stored in shared memory for the next step. If too many triplets are selected, the frame is deemed to complex for online processing and marked for storage.

\begin{algorithm}[t!]
\DontPrintSemicolon
    \tcp{Distribute across threads}
    \ForPar{$hit_0, hit_1, hit_2 \leftarrow cuts$}{
        $3hit\_track \leftarrow fit\_helix(hit_0, hit_1, hit_2)$\;
        $hit_3 \leftarrow find\_closest\_layer3\_hit(3hit\_track)$\;
        $full\_track \leftarrow fit\_helix(3hit\_track, hit_3)$\;
        \If{$full\_track.\chi^2 < \chi^2_\text{max}$}{
            \If{$num\_tracks < max\_tracks$}{
                $store\_track(full\_track)$\;
                $num\_tracks++$\;
            }
        }
    }
    \caption{Track Reconstruction}
    \label{alg:track}
\end{algorithm}

For the track reconstruction, each thread performs a full reconstruction on a triplet and only those with a low enough $\chi^2$ error are kept, as outlined in \cref{alg:track}.
Again if to many tracks are found, the frame is marked for storage and further processing is skipped. 

\begin{algorithm}[t!]
\DontPrintSemicolon
    \tcp{Distribute across threads}
    \ForPar{$e^+_0, e^+_1, e^- \leftarrow tracks$} {
        \uIf{$test\_E_{tot}(e^+_0, e^+_1, e^-)$} {
            store\_track\_comb($e^+_0, e^+_1, e^-$)\;
            $num\_track\_combs++$\:
        }
    }
    \_\_syncthreads()\;
    \uIf{$num\_track\_combs > max\_tracks$} {
        return\;
    }
    \tcp{Distribute across threads}
    \ForPar{$e^+_0, e^+_1, e^- \leftarrow track\_combs$}{
        $intersections \leftarrow get\_intersections(e^+_0, e^+_1, e^-$\;
        $vertex \leftarrow calc\_vertex\_estimate(intersections)$\;
        \If{$vertex\_estimate.\chi^2 > \chi^2_\text{max}$ or keep\_frame}{
            continue\;
        }
        \If{$vertex\_estimate.target\_dist() < target\_dist_\text{max}$ or keep\_frame} {
            continue\;
        }
        \If{$vertex\_estimate.total\_momentum(e^+_0, e^+_1, e^-) > momentum_\text{max} $ or keep\_frame} {
            continue\;
        }
        $keep\_frame \leftarrow true$\;
    }
    \caption{Vertex Fit}
    \label{alg:vertex}
\end{algorithm}
Finally the vertex fit is split into two parts, as seen in \cref{alg:vertex}. 
The first part performs the preliminary check for the track triplet's total energy. 
Those meeting the requirements are stored in shared memory.
When too many track triplets are found, the frame is marked for storage.
Otherwise part two begins, where each track triplet is processed, with one thread processing a chunk of triplets.
Since the vertex reconstruction is computationally expensive and only one possible vertex needs to be found, the algorithm is split into small parts.
At the end of each part, a thread checks if a vertex was found by another thread, in which case all further processing is skipped and the frame is marked for storage.

To allow for asynchronous computing, while new hit data is collected by the FPGA and transferred to the GPU, CUDA streams are used to run the implementation. 
Each stream works on one chunk of frames at a time, therefore the chunk size and number of streams used for the final integration are correlated and still need to be determined. 
For our implementation we divided the GPUs memory in equal chunks for the streams.

\input{chapter/implementation/testing}

%% file: chapter/implementation/testing.tex
\section{Benchmarks}
\label{ch:benchmark}
In our test setup we compare the performance of two different GPUs: a NVidia Geforce GTX 1080Ti and a NVidia Geforce RTX 2080Ti. 
Since the detector setup is not complete yet, the algorithm is tested in an isolated environment using a single computer and simulation data provided by the Mu3e simulation framework built on Geant4.

It was proposed by ~\cite{Bruch2017a} that twelve NVIDIA Geforce GTX 1080Ti are enough to perform the Online Event Selection, building our base line.
For the goal of running on twelve machines, one has to evaluate over $1.302 \cdot 10^6$ frames per second, with $64\,$ns long frames.

\subsection{Efficiency}
Before testing the performance of our implementation, we take a look at its efficiency.
The simulation is set up so that exactly one signal event per frame is happening.
Our implementation is able to reconstruct over $94\%$ of all true particle tracks, with over $97\%$ of tracks from signal particles. 
We are also correctly identifying over $94\%$ of frames where a signal event happened.

Using only hits from signal particles and performing the track and vertex reconstruction on these, results in correctly classifying $\sim 98\%$ of signal frames, similar to the offline test results in~\cite{Bruch2017a}.
The disparity in efficiency is explained by the inaccuracy in track reconstruction, not finding all signal tracks. 
Future versions of this algorithm may use more layers for reconstruction, increasing the track reconstructions performance, possibly resulting in an improved signal frame detection rate.

\subsection{Different Muon Rates}
\begin{figure}[t!]
	\input{chapter/implementation/figures/muon-rates}
	\label{fig:bench-rates}
\end{figure}
Next we test the impact of different muon rates on our implementation.
Our target muon rate for the Mu3e experiment phase I is $1 \cdot 10^8\mu$/s, but higher rates are interesting to study for phase II of the project.

Our implementation processes over $1.4 \cdot 10^6$ frames per second on a 1080Ti, as seen in \cref{fig:bench-rates} (left), reaching our performance goal.
For higher rates this performance almost linearly decreases, down to $\sim 7.4 \cdot 10^4$ frames per second for $1 \cdot 10^9\mu$/s, planned as muon rate for phase~II. 
We notice that here about $20$ times higher computational power is needed for phase~II. 
This performance gap could already be partially closed by using newer GPUs as seen in the performance difference between the 1080Ti and 2080Ti. Here the 2080Ti already doubles the performance across all muon rates.

Besides performance requirements, our goal is to reduce the data rate by a factor of at least $100$.
This goal is reached for our target muon rate, with only $\sim 0.4\%$ of frames kept. 
But with higher muon rates the amount of frames kept increases drastically, as seen in \cref{fig:bench-rates} (right).
In the case of $1 \cdot 10^9\mu$/s about $50\%$ of all frames are kept. 
The main reason is the amount of hit triplets overflowing our set limit.
This limit is currently set to fit all relevant data into shared memory and optimized for our target muon rate.
But it could be improved with a trade-off in memory access times, e.g. by moving the storage into global memory, or iteratively processing a chunk of triplets, then performing a track reconstruction on this chunk, before moving on to the next triplet chunk.
This would result in higher computational costs and may not be enough to reach the data rate reduction of over $100$, since in our current experiments already over $2\%$ of frames are kept, due to a possible event vertex being found.
\begin{figure}[t]
	\centering
	\begin{tikzpicture}
		\begin{axis}[
					xticklabels from table={plots/benchmarks/rates.csv}{x},
					xmode=log,
					xtick=data,
					xlabel={}, 
					ylabel=speedup,
					nodes near coords,
					xticklabel style={rotate=45, anchor=north east, inner sep=0},
					ybar,
					bar width=20pt,
					height=1/2.5*\columnwidth, 
					width=8cm,
					enlarge y limits={upper, abs value=1.5em},
				]
			\addplot+[text=black] table [x={x},y expr={\thisrow{1080Ti} / \thisrow{old}}]{plots/benchmarks/rates.csv};
		\end{axis}
	\end{tikzpicture}
	\begin{tikzpicture}
	    
		\begin{axis}[
					legend style={
						at={(0.0,1)},
						anchor=north west
					},
					xticklabels from table={plots/benchmarks/rates.csv}{x},
					xmode=log,
					xtick=data,
					xlabel=muon  rates,
					ylabel="\%" of frames kept,
	          xticklabel style={rotate=45, anchor=north east, inner sep=0},
					cycle list name=exotic,
					ymajorgrids,
					height=1/2.5*\columnwidth, 
					width=8cm,
				]
		\addplot+[only marks] table [x={x},y={new}]{plots/benchmarks/rates-acc-comp.csv};
		\addplot+[only marks] table [x={x},y={old}]{plots/benchmarks/rates-acc-comp.csv};
		\legend{our implementation, reference implementation}
		\end{axis}
	\end{tikzpicture}
	\vspace*{-.3cm}
	\caption{Top: Performance comparison of our implementation with the version in \cite{Bruch2017a}, displayed as speedup. Our implementation has a speedup of over $2$ for lower muon rates, but loses this advantage for higher rates.\\
	Bottom: Frames kept by our and the reference implementation. While for a muon rate of $1 \cdot 10^8\mu$/s our implementation keeps twice the frames ($\sim 0.4\%$ vs $\sim 0.2\%$), its the other way around for the highest muon rate.}
\label{fig:bench-comp}
\end{figure}

A previous implementation of the Event Selection and Vertex Reconstruction was presented in~\cite{Bruch2017a}.
This reference implementation uses pre-processed hit triplets, performing only the full track and vertex reconstruction on the GPU.
Despite implementing the extra selection step, our implementation beats the reference with a speedup of over $2$ for our target muon rate (\cref{fig:bench-comp}).
With increasing muon rates, the speedup diminishes, even disappears. 
It is important to note here, that the reference implementation stores frames where the amount of possible hit triplet combinations is over $1024$, skipping any processing for this frame. 
In comparison, our implementation always performs Selection Cuts, skipping the remainder of the algorithm only if the amount of hit triplets found is higher than a set value of $768$. 
Combined with different cut values result in varying amount of frames kept between these two implementations.
While for a muon rate of $1 \cdot 10^8\mu$/s the reference implementation keeps only $\sim 0.2\%$ of frames compared to our $\sim 0.4\%$, it skips over $80\%$ of frames for $1 \cdot 10^9\mu$/s. 
This explains the speedup difference, since only about $20\%$ of frames are actually touched for computation, and still $\sim 40\%$ more frames are kept in the end, resulting in a worse reduction rate.

%% file: chapter/implementation/figures/muon-rates.tex
\centering
	\begin{tikzpicture}
	\begin{groupplot} [group style={
				group size=2 by 1, 
				horizontal sep=1.5em, 
				vertical sep=3em,
			},
			width=1/1.9*\columnwidth,
			height=1/2.5*\columnwidth,
			]
		\nextgroupplot[
					legend style={
					    nodes={scale=0.75, transform shape}
					},
					xticklabels from table={plots/benchmarks/rates.csv}{x},
					xmode=log,
					xtick=data,
					xlabel=muon  rates,
					ylabel=frames per second,
	          xticklabel style={rotate=45, anchor=north east, inner sep=0},
					cycle list name=exotic,
					ymajorgrids
				]
		\addplot+[only marks] table [x={x},y={2080Ti}]{plots/benchmarks/rates.csv};
		\addplot+[only marks] table [x={x},y={1080Ti}]{plots/benchmarks/rates.csv};
		\legend{2080Ti, 1080Ti}
		\nextgroupplot[
					legend style={
						at={(0.0,1)},
						anchor=north west,
						nodes={scale=0.75, transform shape}
					},
					xticklabels from table={plots/benchmarks/rates-acc-comp.csv}{x},
					xmode=log,
					xtick=data,
					xlabel=muon  rates,
					yticklabel pos=right,
					ylabel near ticks,
					ylabel=$\%$ of frames kept,
	          xticklabel style={rotate=45, anchor=north east, inner sep=0},
					cycle list name=exotic,
					ymajorgrids,
					ybar stacked,
					nodes near coords,
					nodes near coords align=west,
				]   
		\addlegendimage{empty legend}

		\addplot+[text opacity=0] table [x={x},y={triplet}]{plots/benchmarks/rates-acc-new.csv};
		\addplot+[text opacity=0] table [x={x},y={tracks}]{plots/benchmarks/rates-acc-new.csv};
		\addplot+[text opacity=0] table [x={x},y={found}]{plots/benchmarks/rates-acc-new.csv};
		\addplot+[text=black, text opacity=1] table [x={x},y={vertex}]{plots/benchmarks/rates-acc-new.csv};

		\addlegendentry{\hspace{-.4cm}reason to keep:}
		\addlegendentry{$\#$ hit triplet}
		\addlegendentry{$\#$ tracks}
		\addlegendentry{vertex found}
	\end{groupplot}
	\end{tikzpicture}
	\vspace*{-.3cm}
	\caption{Measurements for different muon rates. Top shows performance measurements for both test systems. Bottom shows the amount of frames kept using the same parameters for different muon rates.}